\documentstyle{mn}

\newif\ifAMStwofonts

\ifoldfss
  \ifCUPmtlplainloaded \else
    \NewTextAlphabet{textbfit} {cmbxti10} {}
    \NewTextAlphabet{textbfss} {cmssbx10} {}
    \NewMathAlphabet{mathbfit} {cmbxti10} {} 
    \NewMathAlphabet{mathbfss} {cmssbx10} {} 
  \fi
  \ifAMStwofonts
    \ifCUPmtlplainloaded \else
      \NewSymbolFont{upmath} {eurm10}
      \NewSymbolFont{AMSa} {msam10}
      \NewMathSymbol{\upi}     {0}{upmath}{19}
      \NewMathSymbol{\umu}     {0}{upmath}{16}
      \NewMathSymbol{\upartial}{0}{upmath}{40}
      \NewMathSymbol{\leqslant}{3}{AMSa}{36}
      \NewMathSymbol{\geqslant}{3}{AMSa}{3E}

      \let\leq=\leqslant 
      \let\geq=\geqslant 
    \fi
  \fi
\fi 

\ifnfssone
  \newmathalphabet{\mathit}
  \addtoversion{normal}{\mathit}{cmr}{m}{it}
  \addtoversion{bold}{\mathit}{cmr}{bx}{it}
  \newmathalphabet{\mathbfit} 
  \addtoversion{normal}{\mathbfit}{cmr}{bx}{it}
  \addtoversion{bold}{\mathbfit}{cmr}{bx}{it}
  \newmathalphabet{\mathbfss} 
  \addtoversion{normal}{\mathbfss}{cmss}{bx}{n}
  \addtoversion{bold}{\mathbfss}{cmss}{bx}{n}
  \ifAMStwofonts
    \ifCUPmtlplainloaded \else
      \UseAMStwoboldmath
      \makeatletter
      \new@mathgroup\upmath@group
      \define@mathgroup\mv@normal\upmath@group{eur}{m}{n}
      \define@mathgroup\mv@bold\upmath@group{eur}{b}{n}
      \edef\UPM{\hexnumber\upmath@group}
      \new@mathgroup\amsa@group
      \define@mathgroup\mv@normal\amsa@group{msa}{m}{n}
      \define@mathgroup\mv@bold\amsa@group{msa}{m}{n}
      \edef\AMSa{\hexnumber\amsa@group}
      \makeatother
      \mathchardef\upi="0\UPM19
      \mathchardef\umu="0\UPM16
      \mathchardef\upartial="0\UPM40
      \mathchardef\leqslant="3\AMSa36
      \mathchardef\geqslant="3\AMSa3E

      \let\leq=\leqslant 
      \let\geq=\geqslant 
    \fi
  \fi
\fi 

\ifnfsstwo
  \DeclareMathAlphabet{\mathbfit}{OT1}{cmr}{bx}{it}
  \SetMathAlphabet\mathbfit{bold}{OT1}{cmr}{bx}{it}
  \DeclareMathAlphabet{\mathbfss}{OT1}{cmss}{bx}{n}
  \SetMathAlphabet\mathbfss{bold}{OT1}{cmss}{bx}{n}
  \ifAMStwofonts
    \ifCUPmtlplainloaded \else
      \DeclareSymbolFont{UPM}{U}{eur}{m}{n}
      \SetSymbolFont{UPM}{bold}{U}{eur}{b}{n}
      \DeclareSymbolFont{AMSa}{U}{msa}{m}{n}
      \DeclareMathSymbol{\upi}{0}{UPM}{"19}
      \DeclareMathSymbol{\umu}{0}{UPM}{"16}
      \DeclareMathSymbol{\upartial}{0}{UPM}{"40}
      \DeclareMathSymbol{\leqslant}{3}{AMSa}{"36}
      \DeclareMathSymbol{\geqslant}{3}{AMSa}{"3E}

      \let\leq=\leqslant 
      \let\geq=\geqslant 
    \fi
  \fi
\fi 

\ifCUPmtlplainloaded \else
  \ifAMStwofonts \else 
    \def\upi{\pi}
    \def\umu{\mu}
    \def\upartial{\partial}
  \fi
\fi

\title{Gas properties of H{\Large {\bf II}} and Starburst galaxies: relation with the stellar population}
\author[D. Raimann et al.]
  {D.~Raimann$^1$, T.~Storchi-Bergmann$^1$, E.~Bica$^1$, J.~Melnick$^2$
  \newauthor
  and H.~Schmitt$^3$ \thanks{E-mail: raimann@if.ufrgs.br; bica@if.ufrgs.br; thaisa@if.ufrgs.br; jmelnick@eso.org; schmitt@stsci.edu} \\
  $^1$ Universidade Federal do Rio Grande do Sul, IF, CP15051, Porto Alegre 91501-970, RS, Brazil \\
  $^2$ European Southern Observatory, Casilla 19001, Santiago 19, Chile \\
  $^3$ Space Telescope Institute, 3700 San Martin Drive, Baltimore, MD21218, USA \\ }

\date{Accepted  
      Received 
      in original form}

\pagerange{\pageref{firstpage}--\pageref{lastpage}}
\pubyear{1999}

\begin{document}

\maketitle

\title{Gas properties of H{\normalsize {\it II}} and Starburst galaxies: relation with stellar population}

\label{firstpage}

\begin{abstract}
We study the gas emission of galaxies with active star formation, consisting mostly of H{\sevensize II} galaxies and Starbursts, as well as some Seyfert 2's, and determine chemical and physical parameters. The data consist of 19 high signal-to-noise ratio optical templates, a result of grouping 185 emission line galaxy spectra. Underlying stellar population models (Raimann et al., 1999) were subtracted from the templates in order to isolate the pure emission component. 

We analyse the distribution of these improved signal-to-noise ratio emission spectra in diagnostic diagrams
and find that the H{\sevensize II} templates show a smaller spread in $log$([O{\sevensize III}]/H$\beta$) values than the individual galaxies, apparently due to the population subtraction and better signal-to-noise ratio. We thus suggest the template sequence as a fiducial observational locus for H{\sevensize II} galaxies which can be used as reference for models. The sequence of line ratios presented by the H{\sevensize II} galaxies in the diagram $log($[O{\sevensize III}]$\lambda5007$/H$\beta) \times log($[N{\sevensize II}]$\lambda$6584/H$\alpha)$ is primarily due to the gas metallicity, of which the $log($[N{\sevensize II}]/H$\alpha)$ ratio is a direct estimator. We also study the properties of the Starburst galaxies and those intermediate between H{\sevensize II} and Starburst galaxies, which are more metal rich and sit on more massive galaxies. 

We discuss the present results in the frame of a recently proposed equivalent width diagnostic diagram for emission line galaxies (Rola et al., 1997) and conclude that the observed ranges in W([O{\sevensize II}])/W(H$\beta)$ and W(H$\beta)$ are mostly due to the non-ionizing stellar population contribution. W(H$\beta)$ can be used as an estimator of this contribution to the continuum, and briefly discuss implications to the cosmological use of H{\sevensize II} galaxies.

\end{abstract}


\section{Introduction}
The investigation of nearby star-forming galaxies plays an important role in the interpretation of the ever increasing data on distant galaxies, as the so-called Lyman-break galaxies seem to be well described by local Starburst galaxies (Meurer et al., 1999). Melnick et al. (1999) propose that H{\sevensize II} galaxies can be used as distance estimators over a wide range of redshifts.

H{\sevensize II} galaxies are among the less luminous star-forming galaxies. Their emission-line spectra and relatively low gaseous metallicities ($\approx$ 0.1 to 0.25 solar, e.g. Pe\~{n}a et al., 1991) would be consistent with the idea that they are very young galaxies undergoing their first episodes of star formation. On the other hand, in a recent study, Schulte-Ladbeck and Crone (1998) have concluded that in the blue compact dwarf galaxy VII Zw 403 there are older stellar population components.

In a previous study (Raimann et al., 1999, hereafter Paper I) we have investigated the stellar population properties of a sample dominated by nearby H{\sevensize II} galaxies, but including also Starburst and Seyfert 2 galaxies for comparison purposes. We have considered their continuum and emission/absorption line properties, grouped them into high signal-to-noise templates, and performed stellar population syntheses using a base of stellar cluster spectra (Bica, 1988; Bica and Alloin, 1986). We concluded that most H{\sevensize II} galaxies present important flux contributions from populations as old as 500 Myr.

In the present paper, we investigate the gaseous properties of the templates obtained in Paper I, after subtraction of the underlying stellar population. We use the emission-line fluxes to classify the spectra according to their excitation characteristics in diagnostic diagrams, to calculate the gaseous abundance and to obtain the age of the last generation of (ionizing) stars. The role of the underlying stellar population, including its effect on the interpretation of data from distant star-forming galaxies and its relation to the emission line properties, is also explored.

We present the data set in section 2. The diagnostic diagrams are discussed in section 3. The gas metallicity, the age of the ionizing stellar population and relation between the metallicity and the non-ionizing stellar population are studied in section 4. Possible uses of emission-line galaxies for cosmological studies are discussed in section 5. The concluding remarks are given in section 6.

\section{The sample}

\begin{figure}
\vspace{22.5cm}
\caption{Representative spectral groups: (a) blue H{\sevensize II}, (b) H{\sevensize II} with Balmer jump in absorption, (c) Seyfert 2. In the bottom panels we present the dereddened spectra and the corresponding synthesized one (Paper I), while in the top panels we present the difference between the two.}
\label{groups}
\begin{tabular}{c}
\includegraphics{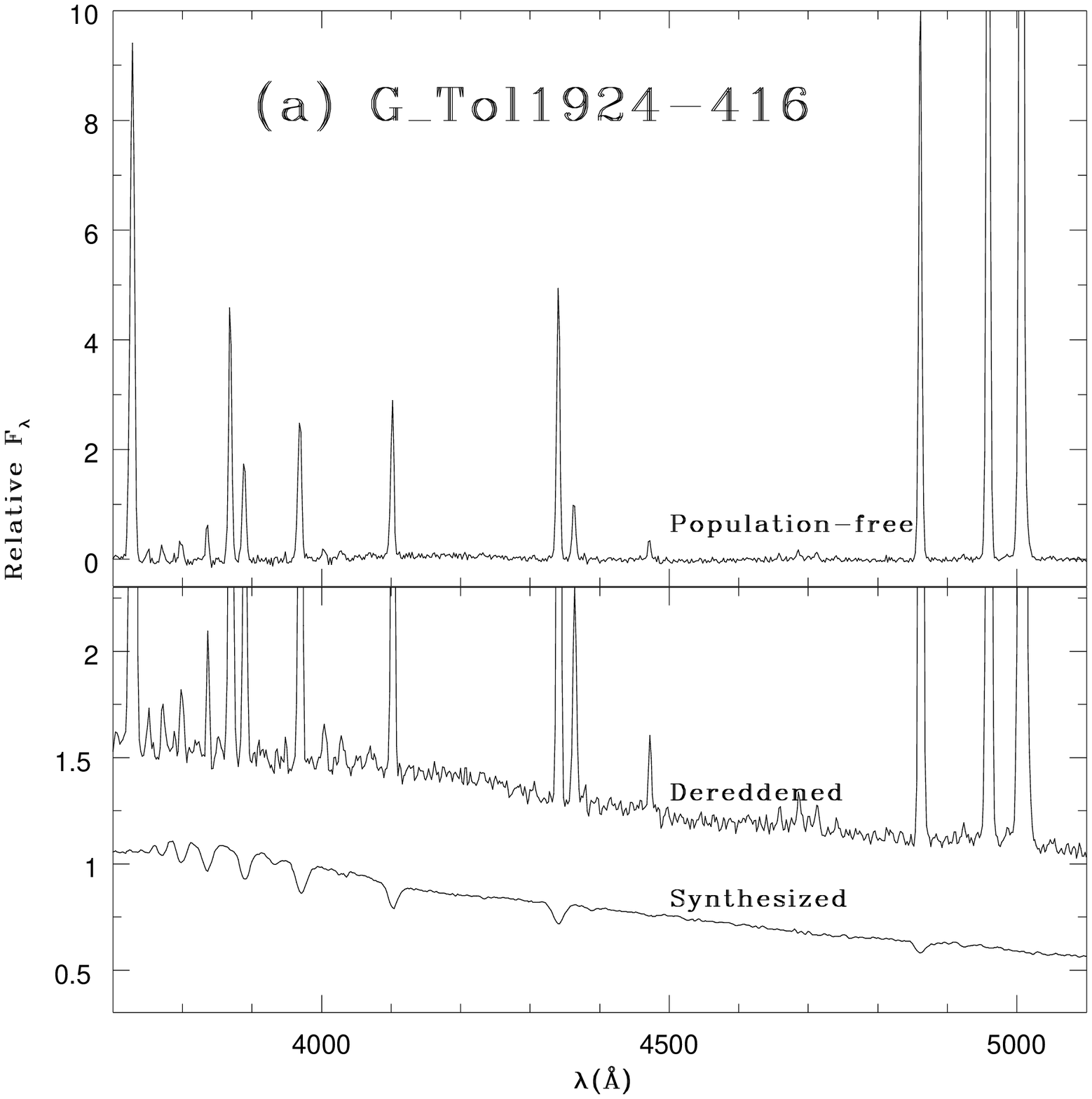} \\
\includegraphics{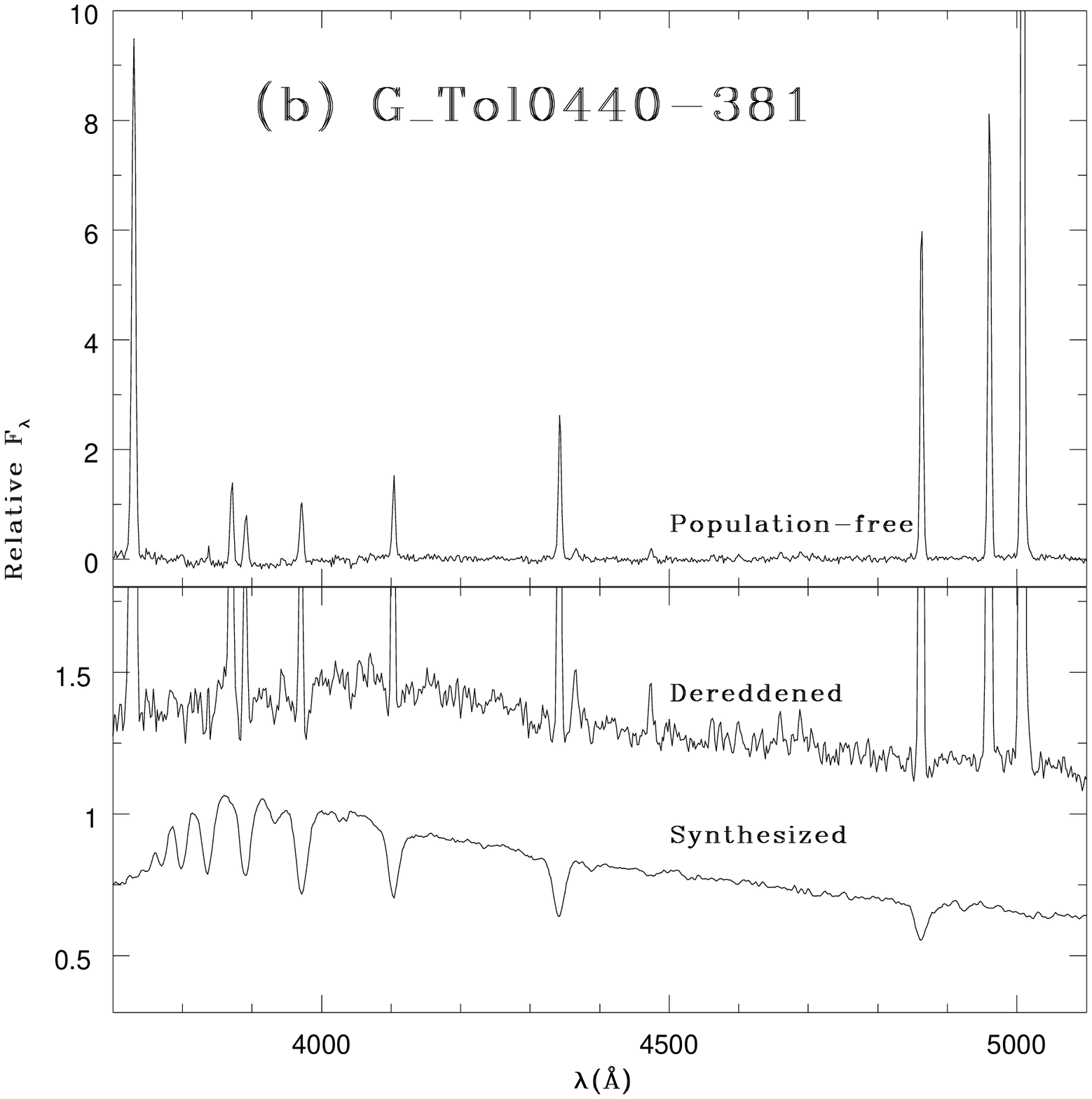} \\
\includegraphics{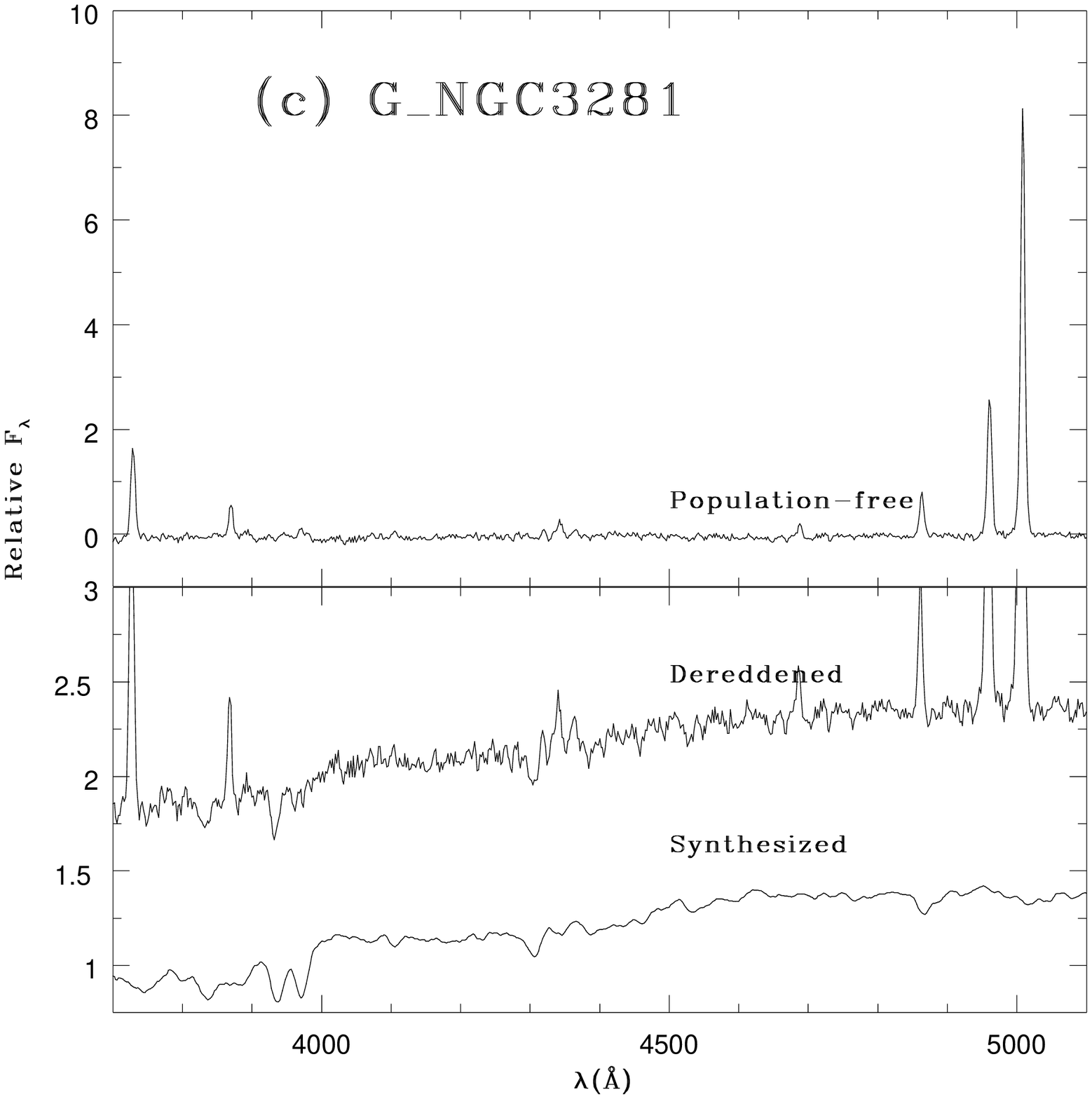} 
\end{tabular}
\end{figure}

The original sample consists of 185 emission-line galaxy spectra obtained by Terlevich and collaborators, most of them (156) discussed individually in Terlevich et al. (1991). The average aperture used in the observations corresponds to $\approx 1.1 \times 1.9$ kpc. More details can be found in Paper I, where we have grouped the spectra according to their continuum, absorption and emission line properties, in order to obtain improved signal-to-noise (S/N) ratio spectra. This strategy was necessary to constrain the stellar population syntheses, mainly for the H{\sevensize II} groups. The resulting spectral groups are listed in Table \ref{grupos} ordered from bluer spectra at the top to redder at the bottom of the table. The groups are named after the member galaxy with the best S/N ratio. The number of the galaxies in each group is shown in column 2. In two cases, G\_UM448 and G\_NGC1510, there were no other similar spectra, and we have thus considered them as ``groups'' of only one galaxy. The spectral type, according to distribution of emission line spectra of the groups in the Baldwin, Phillips and Terlevich (BPT, 1981) diagnostic diagrams, is given in column 3. In column 4 we list the average absolute magnitudes $<$M$_B$$>$, obtained using the apparent magnitudes and radial velocities from the NASA/IPAC Extragalactic Database (NED)\footnote{The NASA/IPAC Extragalactic Database (NED) is operated by the Jet Propulsion Laboratory, California Institute of Technology, under contract with the National Aeronautics and Space Administration.} (see Paper I), for $H_0 = 75$ km s$^{-1}$ Mpc$^{-1}$, and in columns 5, 6 and 7 we provide the average H$\beta$ emission line luminosities, average dimensions at the galaxy corresponding to the angular aperture and average H$\beta$ equivalent width, respectively. In columns 8 and 9 we list the S/N ratios in the continuum of each template for spectral regions around $\lambda$4200\AA\ and $\lambda$5400\AA.

\begin{table*}
 \centering
 \begin{minipage}{180mm}
 \caption{Average properties of the groups.}
 \label{grupos}
\begin{tabular}{lllllcccc} \hline
Groups          & Number of & Spectral Type & \hspace{5mm}$<$M$_B$$>$       & $<$$log $L(H$\beta)$$>$   & $<$Aperture$>$ & $log($W(H$\beta))$ & S/N & S/N\\ 
                & objects   &                &                   & \hspace{5mm}ergs s$^{-1}$            & kpc $\times$ kpc      & \AA\        & 4200\AA & 5400\AA \\   \hline
G\_Cam1148-2020	& 9         & H{\sevensize II} & -17.81 $\pm$ 1.08 & $40.66^{+0.19}_{-0.34}$ & $ 1.5 \times 1.8 $ & 2.37 & 16 & 11 \\ 
G\_UM461	& 9         & H{\sevensize II} & -14.99 $\pm$ 2.68 & $39.27^{+0.19}_{-0.35}$ & $ 0.2 \times 0.3 $ & 2.33 & 21 & 17 \\
G\_Tol1924-416  & 7         & H{\sevensize II} & -19.96 $\pm$ 0.85 & $41.61^{+0.32}_{-1.15}$ & $ 2.5 \times 3.3 $ & 1.98 & 26 & 17 \\
G\_NGC1487      & 6         & H{\sevensize II} & -17.14 $\pm$ 1.07 & $39.17^{+0.40}_{-0.34}$ & $ 0.2 \times 0.3 $ & 1.43 & 19 & 15 \\
G\_Tol1004-296  & 17        & H{\sevensize II} & -17.21 $\pm$ 1.17 & $39.92^{+0.31}_{-1.38}$ & $ 0.4 \times 0.6 $ & 2.03 & 38 & 29 \\
G\_UM448	& 1         & H{\sevensize II} & -19.46            & 41.16                   & $ 1.8 \times 2.1 $ & 1.84 & 21 & 18 \\ 
G\_Tol0440-381  & 18        & H{\sevensize II} & -18.63 $\pm$ 0.67 & $40.92^{+0.38}_{-0.44}$ & $ 1.5 \times 2.7 $ & 1.68 & 21 & 12 \\ 
G\_UM504	& 15        & H{\sevensize II} & -15.86 $\pm$ 0.70 & $39.32^{+0.22}_{-0.44}$ & $ 0.4 \times 0.8 $ & 1.39 & 23 & 17 \\
G\_UM71		& 36        & H{\sevensize II} & -17.68 $\pm$ 1.19 & $40.17^{+0.26}_{-0.71}$ & $ 1.1 \times 2.1 $ & 1.45 & 23 & 16 \\
G\_NGC1510      & 1         & H{\sevensize II} & -16.93            & 39.07                   & $ 0.3 \times 0.4 $ & 1.19 & 26 & 26 \\
G\_Cam0949-2126 & 9         & H{\sevensize II}/Starburst & -19.73 $\pm$ 1.48 & $41.11^{+0.26}_{-0.68}$ & $ 2.5 \times 3.8 $ & 1.47 & 21 & 13 \\
G\_Mrk711       & 5         & H{\sevensize II}/Starburst & -19.81 $\pm$ 0.81 & $41.22^{+0.21}_{-0.48}$ & $ 1.4 \times 2.1 $ & 1.47 & 40 & 26 \\
G\_UM140	& 15        & H{\sevensize II}/Starburst & -18.09 $\pm$ 0.85 & $39.73^{+0.17}_{-0.28}$ & $ 0.8 \times 1.5 $ & 1.07 & 19 & 14 \\
G\_NGC3089      & 11        & H{\sevensize II}/Starburst & -19.29 $\pm$ 1.02 & $40.51^{+0.23}_{-0.54}$ & $ 1.1 \times 2.1 $ & 1.21 & 22 & 16 \\ 
G\_Mrk710       & 2         & Starburst & -18.47            & $39.57^{+0.32}_{-1.27}$ & $ 0.2 \times 0.4 $ & 1.62 & 20 & 19 \\ 
G\_UM477	& 8         & Starburst & -20.11 $\pm$ 0.94 & $40.22^{+0.34}_{-0.77}$ & $ 0.7 \times 1.4 $ & 1.10 & 24 & 26 \\
G\_UM103	& 3         & Seyfert 2 & -19.72 $\pm$ 0.39 & $40.46^{+0.25}_{-0.58}$ & $ 1.8 \times 3.7 $ & 1.18 & 13 & 11 \\
G\_NGC4507      & 7         & Seyfert 2 & -19.77 $\pm$ 0.79 & $40.95^{+0.18}_{-0.31}$ & $ 1.7 \times 3.5 $ & 1.42 & 15 & 19 \\
G\_NGC3281      & 6         & Seyfert 2 & -19.70 $\pm$ 0.96 & $40.06^{+0.30}$         & $ 1.2 \times 2.4 $ & 0.88 & 13 & 18 \\ \hline
\end{tabular}
\end{minipage}
\end{table*}

The spectra of the groups were dereddened according to $E(B-V)_{i}$, the internal reddening affecting the stellar population which was obtained in the synthesis (Paper I). The synthesized stellar population from Paper I was then subtracted from the spectrum of each group. The importance of the subtraction of the underlying stellar population for emission line measurements has been discussed in detail by Bonatto et al. (1989). The main effect is on H$\beta$, with consequences on gas internal reddening determination and calculated emission line ratios involving this line. Three representative spectra are shown in Fig. \ref{groups}. In the top of each panel we show the population-free spectrum and in the bottom the dereddened spectrum previously to stellar population subtraction together with the corresponding synthesized spectrum used in the subtraction. The emission spectra for the remaining groups were presented in Raimann (1998).

The emission-line fluxes of the population-free spectra were then measured and  the corresponding values, relative to  H$\beta$, are shown in Table \ref{Flux}. In the cases of line blending (e.g. H$\alpha+$[N{\sevensize II}]$\lambda\lambda6548,84$ and [S{\sevensize II}]$\lambda\lambda6717,31$) the profiles were constrained by adjusting Gaussians of the same width. We have then calculated the residual gas reddening assuming a Galactic reddening law (Seaton, 1979), case B recombination, and an intrinsic ratio H$\alpha$/H$\beta=2.9$ (Osterbrock, 1989). Only two groups, G\_NGC1510 and G\_NGC3281, presented significant gas reddenings with $E(B-V)_{gas}=$ 0.24 and 0.32, respectively. The other groups have $E(B-V)_{gas}\leq 0.05$, which are listed in column 2 of Table \ref{result}. 

\begin{table*}
\centering
\begin{minipage}{160mm}
 \caption{Emission line fluxes relative to H$\beta$.}
 \label{Flux}
 \begin{tabular}{lrrrrrrrrrrr} \hline
Group         & [O{\sevensize II}]       & [Ne{\sevensize III}]     &  H$\gamma$     &  [O{\sevensize III}]     &  H$\beta$      & [O{\sevensize III}]      &   [O{\sevensize I}]        & H$\alpha$     & [N{\sevensize II}]       &  [S{\sevensize II}]       &  [S{\sevensize II}] \\
              & $\lambda3727$ & $\lambda3869$ & $\lambda4340$ &  $\lambda4363$ &  $\lambda4861$ & $\lambda5007$ &  $\lambda6300$ & $\lambda6563$ & $\lambda6584$ & $\lambda6717$ &  $\lambda6731$ \\ \hline
G\_Cam1148-2020	& 0.89 & 0.48 & 0.50 & 0.13 & 1.00 & 6.25 & 0.02 & 2.67 & 0.04 & 0.07 & 0.05  \\
G\_UM461	& 0.86 & 0.51 & 0.48 & 0.12 & 1.00 & 6.06 & 0.01 & 2.34 & 0.03 & 0.05 & 0.04  \\
G\_Tol1924-416	& 1.15 & 0.45 & 0.48 & 0.10 & 1.00 & 5.52 & 0.02 & 2.05 & 0.04 & 0.07 & 0.04  \\
G\_NGC1487	& 2.64 & 0.33 & 0.48 & 0.03: & 1.00 & 3.31 & $\leq 0.01$ & 2.64 & 0.21 & 0.24 & 0.16  \\
G\_Tol1004-296	& 1.53 & 0.37 & 0.47 & 0.05 & 1.00 & 4.74 & 0.03 & 2.77 & 0.11 & 0.12 & 0.09  \\
G\_UM448	& 2.95 & 0.22 & 0.51 & 0.02 & 1.00 & 2.85 & 0.05 & 2.69 & 0.32 & 0.21 & 0.17  \\
G\_Tol0440-381	& 1.98 & 0.26 & 0.43 & 0.03 & 1.00 & 4.04 & 0.06 & 2.44 & 0.15 & 0.11 & 0.10  \\
G\_UM504	& 2.48 & 0.22 & 0.34 & 0.03: & 1.00 & 3.64 & $\leq 0.01$ & 2.77 & 0.23 & 0.29 & 0.18  \\
G\_UM71		& 2.19 & 0.24 & 0.45 & 0.02: & 1.00 & 3.75 & $\leq 0.01$ & 2.70 & 0.22 & 0.17 & 0.10  \\
G\_NGC1510	& 1.80 & 0.19 & 0.40 & 0.03: & 1.00 & 3.93 & $\leq 0.01$ & 3.75 & 0.39 & 0.29 & 0.20  \\
G\_Cam0949-2126 & 2.19 & 0.09 & 0.38 & $\leq 0.01$ & 1.00 & 1.83 & $\leq 0.01$ & 2.58 & 0.42 & 0.16 & 0.21  \\ 
G\_Mrk711	& 2.26 & 0.14 & 0.44 & $\leq 0.01$ & 1.00 & 2.16 & 0.09 & 3.01 & 0.88 & 0.30 & 0.23  \\
G\_UM140	& 2.75 & $\leq 0.01$ & 0.44 & $\leq 0.01$ & 1.00 & 1.35 & $\leq 0.01$ & 2.87 & 0.59 & 0.27 & 0.34  \\
G\_NGC3089	& 2.03 & $\leq 0.01$ & 0.37 & $\leq 0.01$ & 1.00 & 1.09 & $\leq 0.01$ & 3.04 & 0.96 & 0.38 & 0.27  \\
G\_Mrk710 	& 1.20 & $\leq 0.01$ & 0.44 & $\leq 0.01$ & 1.00 & 0.35 & $\leq 0.01$ & 3.07 & 1.12 & 0.28 & 0.26  \\
G\_UM477	& 1.05 & $\leq 0.01$ & 0.43 & $\leq 0.01$ & 1.00 & 0.26 & $\leq 0.01$ & 3.06 & 1.46 & 0.35 & 0.33  \\
G\_UM103	& 2.52 & 0.63 & 0.34 & 0.11 & 1.00 & 7.86 & 0.27 & 2.55 & 1.22 & $\leq 0.01$ & 0.44  \\
G\_NGC4507	& 1.76 & 0.83 & 0.49 & 0.22 & 1.00 & 10.2 & 0.37 & 2.94 & 1.75 & 0.49 & 0.51  \\
G\_NGC3281	& 2.24 & 0.68 & 0.51 & 0.22: & 1.00 & 10.4 & 0.41 & 4.08 & 4.04 & 1.00 & 0.96  \\ \hline
\end{tabular}
\medskip
Note: The colon symbol indicates a large uncertainty.
\end{minipage}
\end{table*}

\section{BPT Diagnostic diagrams}

\begin{figure}
\vspace{8cm}
\includegraphics{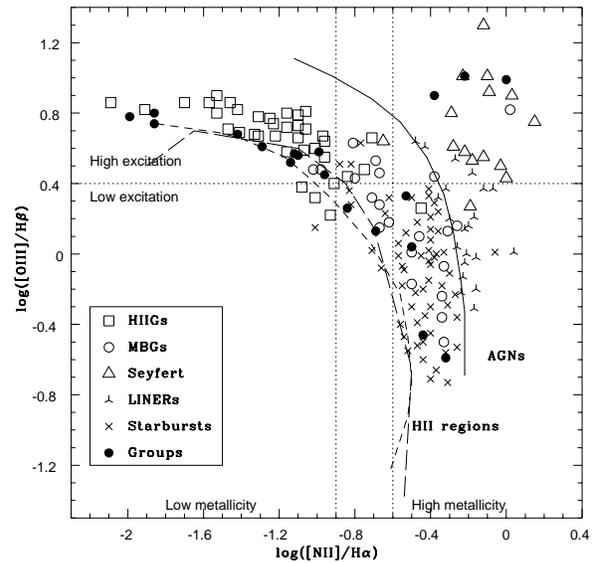}
\caption{BPT diagnostic diagram $log($[O{\sevensize III}]$\lambda$5007/H$\beta)$ $\times$  $log($[N{\sevensize II}]$\lambda$6584/H$\alpha)$, adapted from Coziol (1996) showing the location of the 19 groups (filled circles) as compared to other emission line galaxies. The continuous line to the right is an empirical borderline separating H{\sevensize II} regions from AGNs. The short-dashed curve represents H{\sevensize II} region models of Evans and Dopita (1985) and the long-dashed curve, H{\sevensize II} region models of McCall et al. (1985).}
\label{Diag}
\end{figure}

\begin{figure*}
\vspace{8cm}
\caption{BPT diagnostic diagrams (a) $log($[O{\sevensize III}]$\lambda$5007/H$\beta)$ $\times$ $log($[S{\sevensize II}]$\lambda\lambda$6717,6731/H$\alpha)$ and (b) $log($[O{\sevensize III}]$\lambda$5007/H$\beta)$ $\times$ $log($[O{\sevensize I}]$\lambda$6300/H$\alpha)$. Symbols and curves as in Fig. \ref{Diag}. In diagram (b) arrows indicate upper limits.}
\label{Diag2}
\begin{tabular}{cc}
\includegraphics{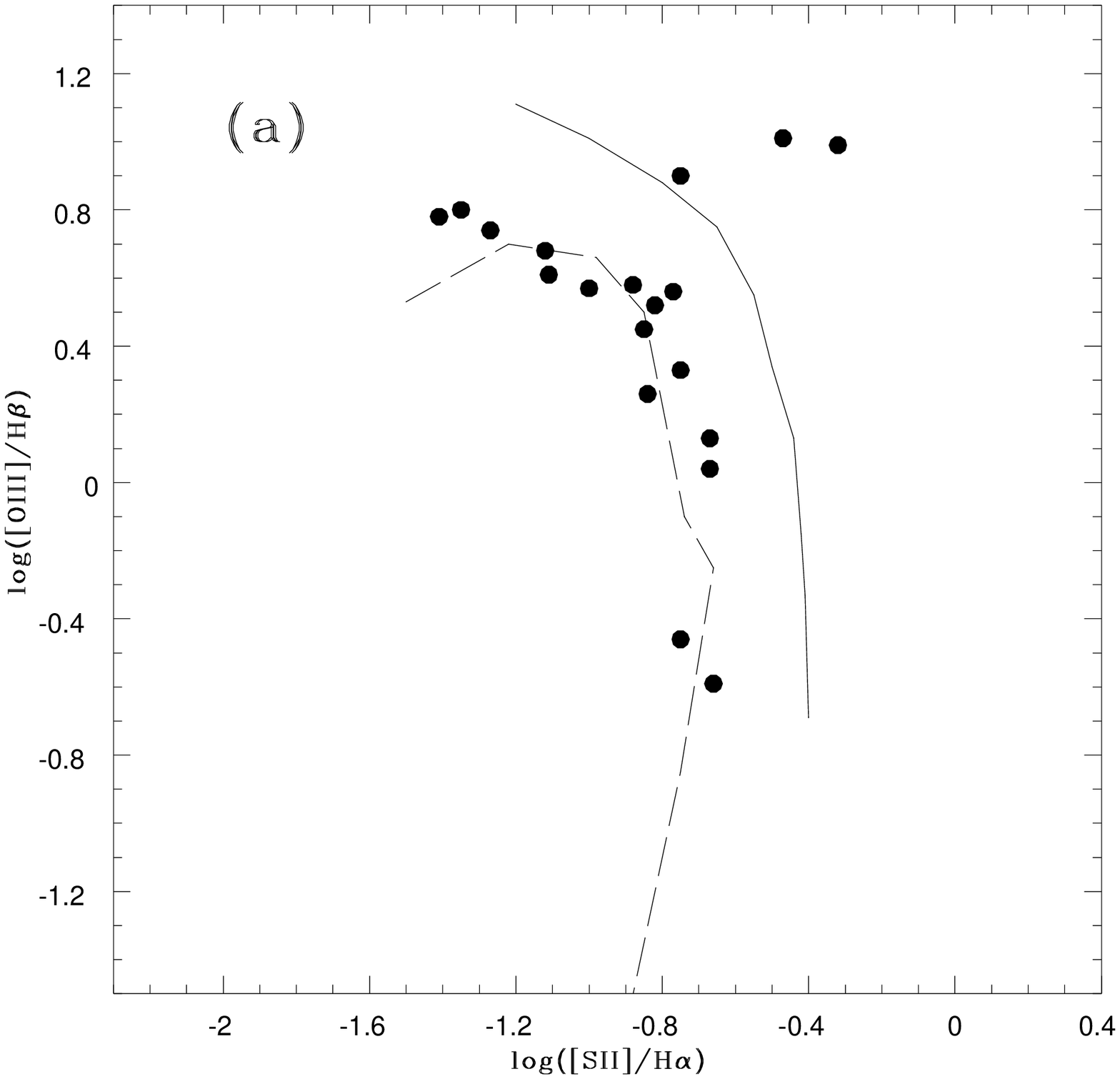} &
\includegraphics{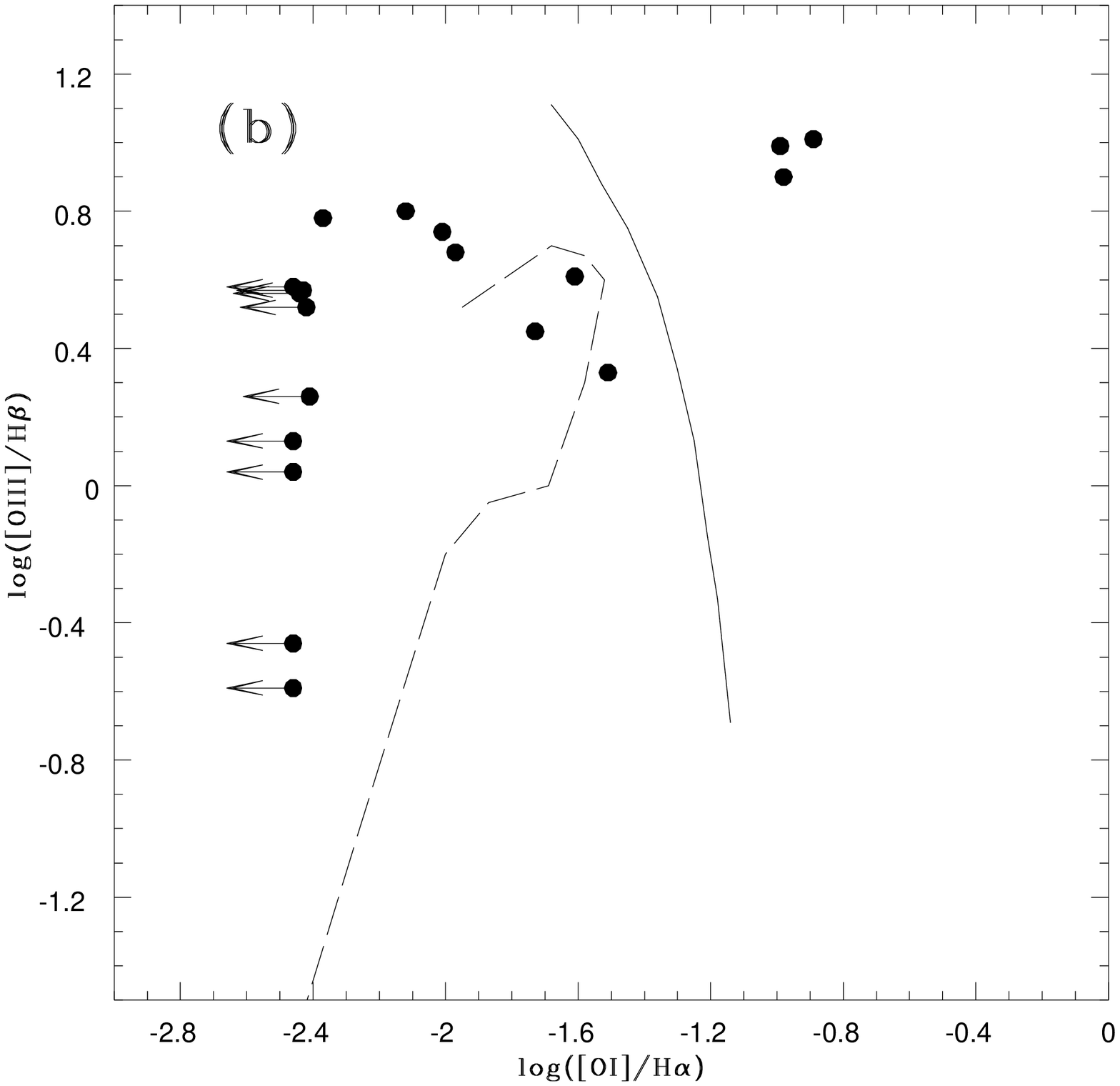}
\end{tabular}
\end{figure*}

The reddening-corrected emission-line fluxes were used to locate the groups in the BPT diagnostic diagrams $log($[O{\sevensize III}]$\lambda5007$/H$\beta)$ $\times$ $log($[N{\sevensize II}]$\lambda$6584/H$\alpha)$, $log($[O{\sevensize III}]$\lambda5007$/H$\beta)$ $\times$ $log($[S{\sevensize II}]$\lambda\lambda$6717,6731/H$\alpha)$ and $log($[O{\sevensize III}]$\lambda5007$/H$\beta)$ $\times$ $log($[O{\sevensize I}]$\lambda$6300/H$\alpha)$ (Baldwin et al., 1981; Veilleux and Osterbrock, 1987), shown in Figs. \ref{Diag} and \ref{Diag2}.

Fig. \ref{Diag} contains the present galaxy groups data, together with those of Starburst Nuclear (therein referred to as MGB), Seyfert, LINER and H{\sevensize II} galaxies from Coziol (1996). In this diagram the groups G\_UM103, G\_NGC4507 and G\_NGC3281 are located in the region occupied by Seyfert 2  galaxies while G\_UM477 and G\_Mrk710 are in the Nuclear Starburst region. The groups G\_Mrk711, G\_Cam0949-2126, G\_UM140 and G\_NGC3089 are located in a region intermediate between those of H{\sevensize II} and Nuclear Starbursts. The remaining groups -- the majority -- are located in the H{\sevensize II} galaxies' region. The classification is shown in column 3 of Table \ref{grupos}. It can be observed that the H{\sevensize II} galaxy groups are located closer to the H{\sevensize II} region loci as defined by the dashed lines in this diagram than the individual H{\sevensize II} galaxies of Coziol's (1996) sample.

Fig. \ref{Diag2} contains only our groups data, together with the H{\sevensize II} models of McCall et al. (1985). In Fig. \ref{Diag2}a the H{\sevensize II} galaxy groups are also located close to the H{\sevensize II} region loci as defined by the dashed line. 

\begin{figure*}
\vspace{8cm}
\caption{BPT diagnostic diagrams (a) $log($[O{\sevensize III}]$\lambda$5007/H$\beta)$ $\times$ $log($[N{\sevensize II}]$\lambda$6584/H$\alpha)$ and (b) $log($[O{\sevensize III}]$\lambda$5007/H$\beta)$ $\times$ $log($[S{\sevensize II}]$\lambda\lambda$6717,6731/H$\alpha)$      comparing the groups (filled circles) with the individual galaxies from Terlevich et al. (1991, crosses).}
\label{Terl}
\begin{tabular}{cc}
\includegraphics{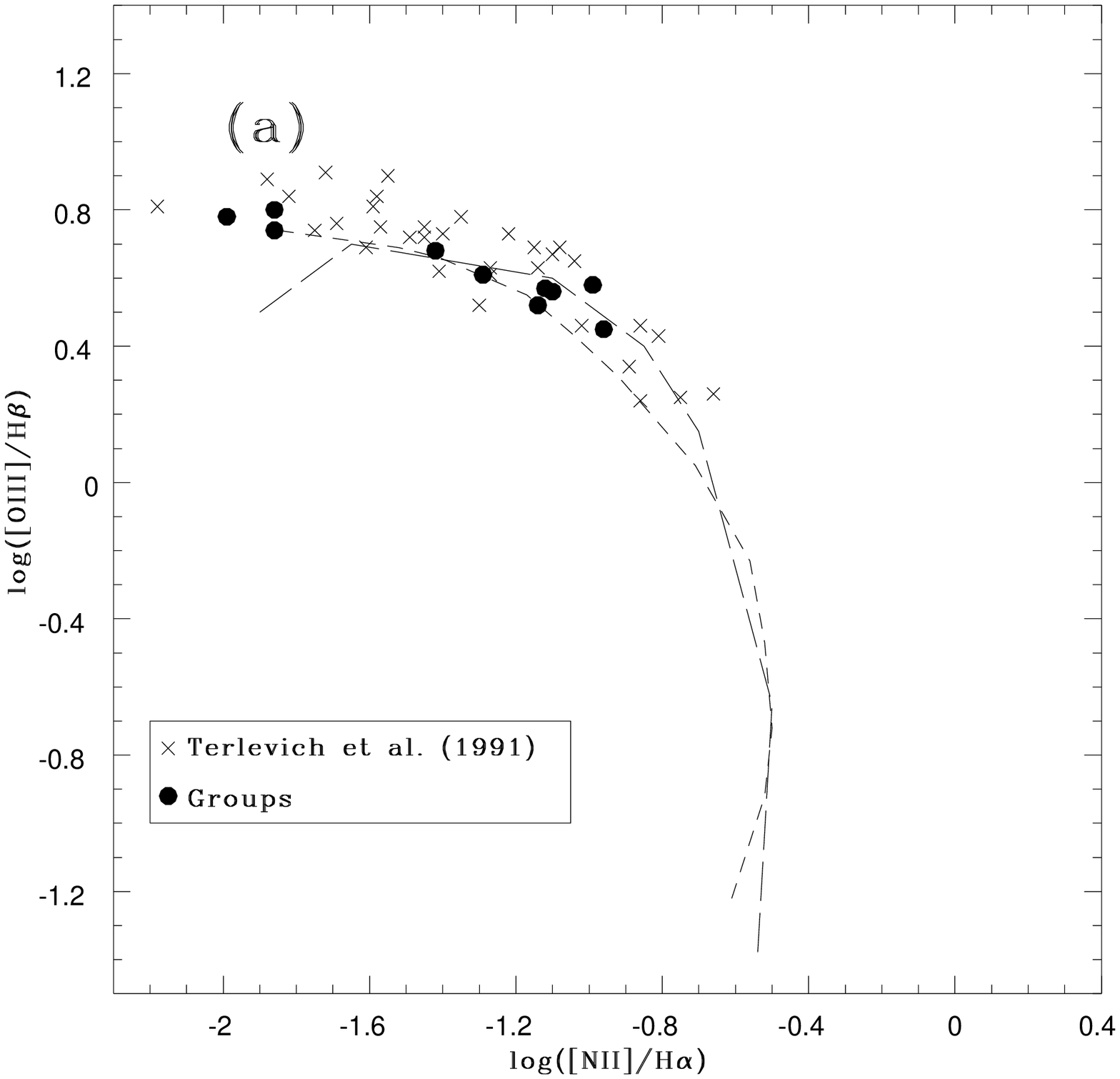} &
\includegraphics{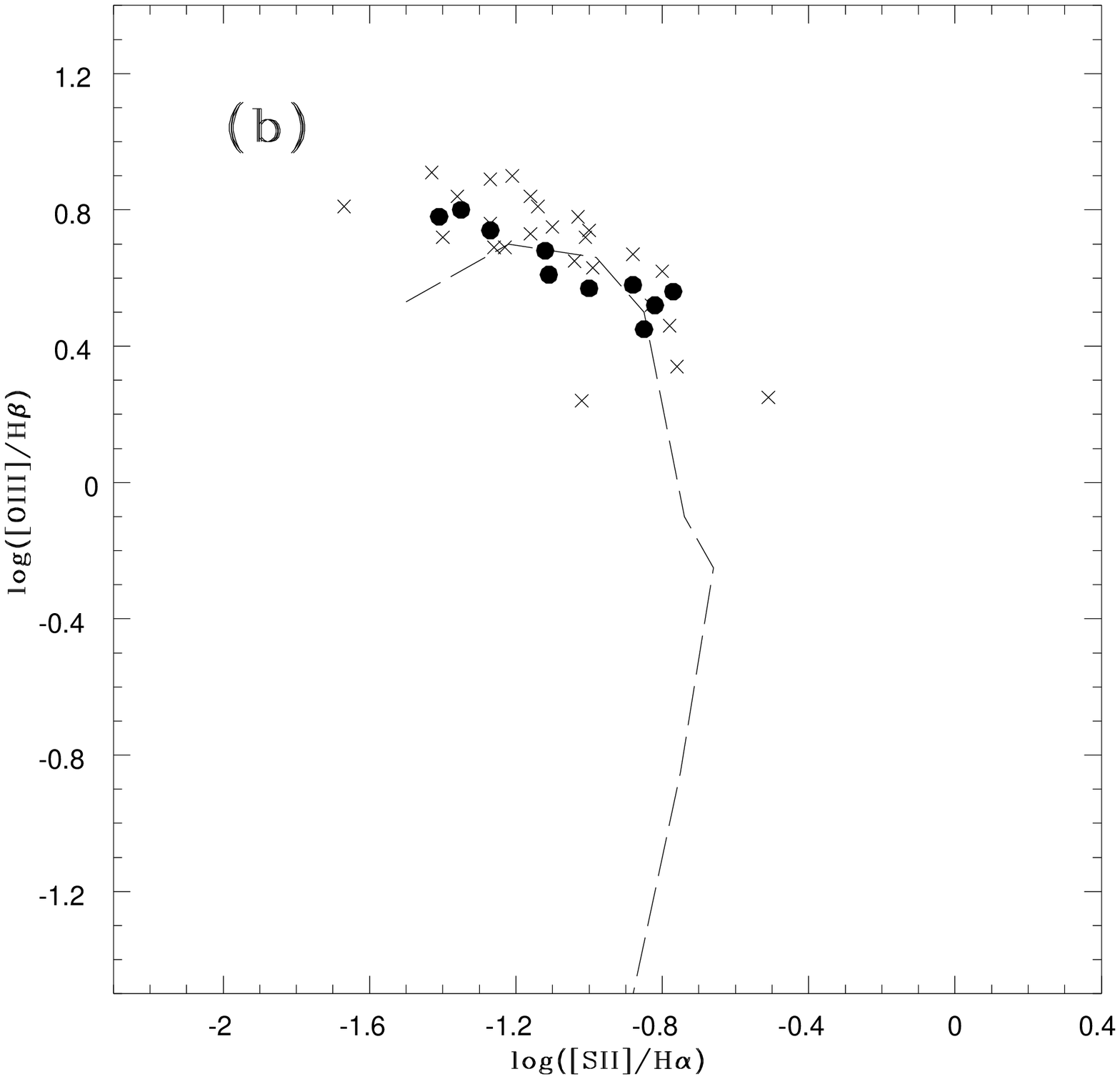}
\end{tabular}
\end{figure*}

In order to check if the groups are indeed closer to the theoretical loci than the individual HII galaxies, we have plotted in Fig. \ref{Terl} the groups data together with the original data for each individual galaxy (Terlevich et al., 1991). It can be observed that the effect is still present. Two factors contribute to this. The subtraction of the stellar population which increases the H$\beta$ emission (reducing $log($[O{\sevensize III}]/H$\beta)$) and the grouping of the individual spectra which produces higher S/N spectra. In H{\sevensize II} galaxies with significant contribution of $t \geq 50$ Myr stellar populations the first effect is particularly important, because these components have strong Balmer absorption lines.

We conclude that with spectra of high S/N ratio and corrected for the stellar population absorptions the H{\sevensize II} galaxies in the BPT diagnostic diagrams $log($[O{\sevensize III}]$\lambda5007$/H$\beta)$ $\times$ $log($[N{\sevensize II}]$\lambda$6584/H$\alpha)$ and $log($[O{\sevensize III}]$\lambda5007$/H$\beta)$ $\times$ $log($[S{\sevensize II}]$\lambda\lambda$6717,6731/H$\alpha)$ get closer to the theoretical H{\sevensize II} region loci presented by Evans and Dopita (1985) and McCall et al (1985). Previous works have proposed models with very high temperatures to cover the region occupied by H{\sevensize II} galaxies (Tresse et al., 1996 and Rola et al., 1997). Our results suggest that this is not necessary and that H{\sevensize II} galaxies behave similarly to H{\sevensize II} regions. We therefore suggest this observational locus as reference for theoretical models of H{\sevensize II} galaxies in the BPT diagrams. 

In the diagnostic diagram $log($[O{\sevensize III}]$\lambda$5007/H$\beta)$ $\times$ $log($[O{\sevensize I}]$\lambda$6300/H$\alpha)$ (Fig. \ref{Diag2}b) the galaxy groups for which it was possible to measure [O{\sevensize I}]$\lambda$6300 are located close to the H{\sevensize II} region loci as defined by the dashed line. However, for many groups -- mainly those with the lowest [O{\sevensize III}]/H$\beta$ -- we have obtained only upper limits for [O{\sevensize I}]$\lambda$6300; these limits suggest sistematically lower values for the galaxies when compared to H{\sevensize II} region models. On the other hand, there are not enough H{\sevensize II} regions data in the literature with low [O{\sevensize III}]/H$\beta$ and with measured [O{\sevensize I}]$\lambda$6300, so these models are difficult to be compared even with H{\sevensize II} region data.

The only star forming group which approaches considerably the AGN region is the H{\sevensize II}/Starburst group G\_Mrk711. It would be interesting to study in detail galaxies in this group (Paper I), because they might contain intrinsically high temperature H{\sevensize II} regions, or alternatively that locus might reflect a composite spectrum, i.e. AGN mixed to H{\sevensize II} region emissions due to aperture effects (Pastoriza et al., 1999 and Storchi-Bergmann, 1991).

\section{Age and metallicity effects}

We now study the gas properties, using the population-free spectra, in order to investigate the gas metallicity and age of the ionizing stellar population, and their relation to the loci occupied by the groups in the BPT diagrams.  

\subsection{Metallicity}

Since the metallicity is very sensitive to gas temperature, which is not uniform in the nebulae, it is important to consider the ionization structure (Garnett, 1992). We have adopted the two-zone model of Campbell et al. (1986), where the emission-line ratio [O{\sevensize III}]$\lambda\lambda4959,5007/\lambda4363$ gives the temperature ($T^{++}$) of the high ionization species (like O$^{++}$ and Ne$^{++}$) and the emission line ratio [N{\sevensize II}]$\lambda\lambda6548,84/\lambda5755$ gives the temperature ($T^+$) of the low ionization species (like O$^+$, N$^+$ and S$^+$). The density is obtained through the emission-line ratio [S{\sevensize II}]$\lambda6717/\lambda6731$.

We were able to measure [O{\sevensize III}]$\lambda4363$ for the H{\sevensize II} galaxies and Seyfert 2's, while for the remaining groups only upper limits could be obtained (column 5 of Table \ref{Flux}). For the latter groups we have used the empirical calibration of Pagel et al. (1979) extrapolated as described in Schmitt et al. (1994) and Storchi-Bergmann et al. (1996) to calculate $T^{++}$. We could not detect [N{\sevensize II}]$\lambda5755$ in any of the spectra. We have thus used the relation of Campbell et al. (1986) to calculate $T^+$. The ionic abundance calculations were performed using a three-level atom model (McCall, 1984).

The resulting $T^{++}$, Ne and ionic abundances are shown in Table \ref{result}. The total oxygen abundance (O/H) was calculated by adding the contributions of O$^0$, O$^+$ and O$^{++}$. The nitrogen abundance was calculated by assuming that N/O=N$^+$/O$^+$, based upon the rough coincidences between the ionization potentials of the two ions (Storchi-Bergmann et al., 1994 and references therein). 

\begin{table*}
\centering
\begin{minipage}{160mm}
 \caption{Analysis results.}
 \label{result}
 \begin{tabular}{lrrrrrrrr} \hline
Group &  $E(B-V)_{gas}$ & Ne(cm$^{-3}$) & $T^{++}$(K) & (N$^+$/H$^+$)%
\footnote{In 12+log units.} & (S$^+$/H$^+)^a$ & 
(O/H)$^{a,}$\footnote{12+log(O/H)$_{\odot}=8.91$ and log(N/O)$_{\odot}=-0.93$.} &  (Ne$^{++}$/H$^+)^a$ & log(N/O)$^b$\\ \hline
G\_Cam1148-2020	& 0.00 & $\approx 100$ & 15475 & 5.52 & 5.10 & 7.87 & 7.04 & -1.44 \\
G\_UM461	& 0.00 & $\approx 100$ & 15039 & 5.29 & 5.01 & 7.89 & 7.11 & -1.70 \\
G\_Tol1924-416	& 0.00 & 300           & 14436 & 5.56 & 5.11 & 7.92 & 7.10 & -1.57 \\
G\_NGC1487	& 0.00 & $\approx 100$ & 11177 & 6.51 & 5.89 & 8.19 & 7.32 & -1.35 \\
G\_Tol1004-296	& 0.00 & $\approx 100$ & 11811 & 6.17 & 5.57 & 8.14 & 7.30 & -1.39 \\
G\_UM448	& 0.00 & 258           & 10294 & 6.75 & 5.94 & 8.32 & 7.29 & -1.28 \\
G\_Tol0440-381	& 0.00 & 330           & 10475 & 6.41 & 5.68 & 8.31 & 7.33 & -1.44 \\
G\_UM504	& 0.00 & $\approx 100$ & 10839 & 6.57 & 5.97 & 8.23 & 7.21 & -1.29 \\
G\_UM71		& 0.00 & 151           & 10082 & 6.62 & 5.57 & 8.32 & 7.36 & -1.28 \\
G\_NGC1510	& 0.24 & $\approx 100$ & 11011 & 6.67 & 5.87 & 8.21 & 7.20 & -1.13 \\
G\_Cam0949-2126	& 0.00 & 1654          & 7776 & 7.15 & 6.27 & 8.63 & 7.45  & -1.26 \\ 
G\_Mrk711	& 0.03 & 141           & 9109 & 7.40 & 6.28 & 8.58 & 7.58  & -0.87 \\
G\_UM140	& 0.00 & 1354          & 7727 & 7.30 & 6.48 & 8.65 &       & -1.20 \\
G\_NGC3089	& 0.04 & $\approx 100$ & 7144 & 7.56 & 6.48 & 8.61 &       & -0.86 \\
G\_Mrk710 	& 0.05 & 350           & 4968 & 8.00 & 6.78 & 8.99 &       & -0.81 \\
G\_UM477	& 0.05 & 330           & 4611 & 8.19 & 6.95 & 9.05 &       & -0.69 \\
G\_UM103	& 0.00 & 3080          & 13475&  &   & 8.69\footnote{Calculated using a different method: the calibration of Storchi Bergmann et al. (1998; see text).} &  & \\
G\_NGC4507	& 0.01 & 1048          & 15797&  &   & 8.92$^c$ &   & \\
G\_NGC3281	& 0.32 & 755           & 17285&  &   & 8.83$^c$ &   & \\ \hline
 \end{tabular}
\end{minipage}
\end{table*}

The ten H{\sevensize II} galaxy groups (G\_NGC1510, G\_UM504, G\_UM71, G\_UM461, G\_Tol1924-416, G\_UM448, G\_NGC1487, G\_Tol1004-296, G\_Tol0440-381 and  G\_Cam1148-2020) span a range in oxygen abundance of $7.87<12+log$(O/H)$<8.32$. With small variations, because of different methodologies and sample, these values agree with those of Pe\~{n}a et al. (1991) and Terlevich et al. (1991). The relatively large metallicities attained by the H{\sevensize II} galaxies are consistent with the result found in Paper I, that most H{\sevensize II} galaxy groups are not single generation, but present previous generations of stars which have enriched the gas. The groups classified as Starbursts -- G\_UM477 and G\_Mrk711 -- have near solar metallicity, while the ones classified as intermediate between H{\sevensize II} and Starburst -- G\_Cam0949-2126, G\_Mrk711, G\_UM140 and G\_NGC3089 -- have somewhat lower metallicity: 12+log(O/H)$\approx$8.6.

It can be noticed in Table \ref{result} that the Seyfert 2 groups G\_UM103, G\_NGC3281 and G\_NGC4507 have very high temperatures when compared with those predicted by photoionization models (Storchi-Bergmann et al., 1990) which make the calculated chemical abundances lower than expected. This result is typical of Seyfert galaxies and is due to density stratification, to the presence of matter-bounded clouds (Binette et al., 1996), collisional de-excitation and a hard ionizing continuum, effects not present in H{\sevensize II} regions (Viegas-Aldrovandi and Gruenwald, 1988), making the above temperature calibration not valid for the Seyferts. Besides, Storchi-Bergmann et al. (1996) have pointed out how critical is the stellar population subtraction to the measurement of the [O{\sevensize III}]$\lambda4363$ in Seyfert's, which usually causes one to overestimate the strength of this line.

We have thus revised the metallicities of the Seyfert 2 groups using the calibrations of Storchi-Bergmann et al. (1998) for the narrow-line region (NLR) of active galaxies which do not depend on [O{\sevensize III}]$\lambda4363$. These calibrations are a function of the emission-line ratios [O{\sevensize III}]$\lambda\lambda$4959,5007/H$\beta$, [N{\sevensize II}]$\lambda\lambda$6548,84/H$\alpha$ and [O{\sevensize II}]$\lambda3727$/[O{\sevensize III}]$\lambda\lambda4959,5007$. We find that the (O/H) abundance is solar for G\_NGC4507 and slightly below solar for G\_UM103 and G\_NGC3281.

\begin{figure}
\vspace{8cm}
\includegraphics{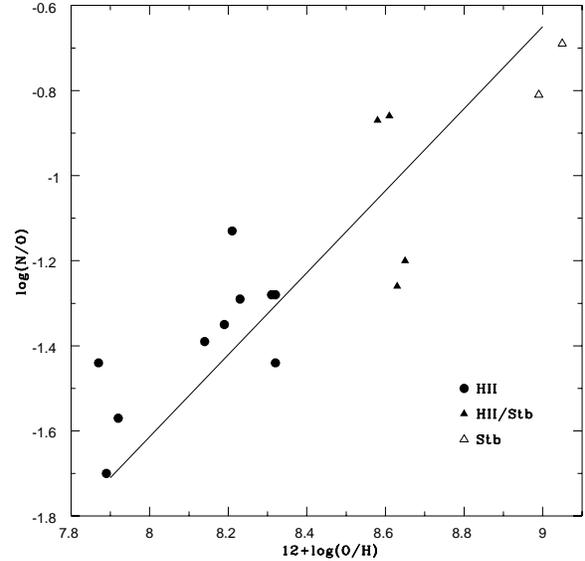}
\caption{$log$(N/O) $\times$ $12+log$(O/H) and linear regression of Storchi-Bergmann et al. (1994). Circles represent H{\sevensize II} galaxies, open triangles Starbursts and filled triangles Intermediate H{\sevensize II}/Starbursts.}
\label{no_oh}
\end{figure}

In Fig. \ref{no_oh} we show $log$(N/O) plotted against $12+log$(O/H) for the H{\sevensize II} galaxies and Starbursts. The distribution of our galaxy groups is consistent with the relation found by Storchi-Bergmann et al. (1994) for star-forming galaxies covering 8.3 $<$ 12+log(O/H) $<$ 9.4 (solid line in this figure). This relation is predicted by a simple model of galactic chemical evolution with instantaneous recycling, in which nitrogen has a secondary origin and oxygen has primary origin (Storchi-Bergmann et al., 1994 and references therein). Vila-Costas and Edmunds (1993) have shown that the secondary behavior dominates for 12+$log$(O/H) $>$ 8.2, while for lower abundances nitrogen has mainly a primary origin [log(O/N) is approximately constant]. We have too few galaxies in the low abundance end to reach any conclusion there, but we can say that the data are consistent also with the results of Vila-Costas and Edmunds, as all data points in the low metallicity end are located above the line denoting secondary behavior.     

\begin{figure}
\vspace{8cm}
\includegraphics{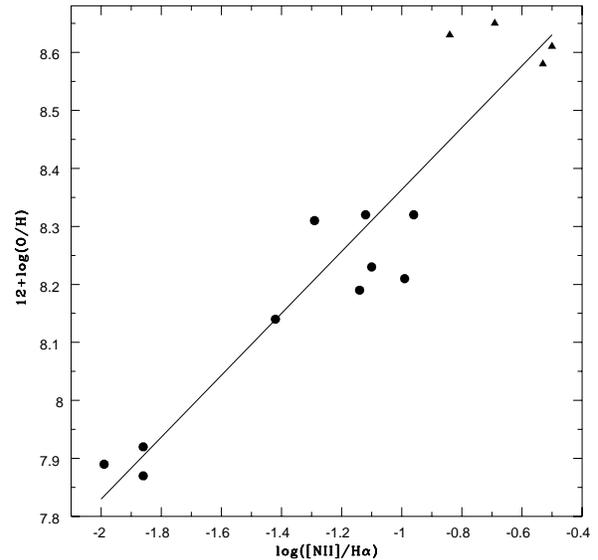}
\caption{$12+log$(O/H) $\times log($[N{\sevensize II}]/H$\alpha)$ and a linear regression to the data. Symbols as in Fig. \ref{no_oh}.}
\label{nii}
\end{figure}

Finally, we notice that the BPT sequence of H{\sevensize II} galaxies is also a sequence of [N{\sevensize II}]/H$\alpha$ values. We have thus plotted in Fig. \ref{nii}, $12+log$(O/H) $\times log($[N{\sevensize II}]/H$\alpha)$ for the H{\sevensize II} and Intermediate H{\sevensize II}/Starburst groups and conclude that they are very well correlated, with a Spearman rank correlation coefficient of 0.90. $Log($[N{\sevensize II}]/H$\alpha)$ can thus be used as a metallicity index for these galaxies. A linear regression to the data in Fig. \ref{nii} gives:

\[ 12+log(O/H)=8.89(\pm0.07)+0.53(\pm0.06) \times log([N{\sevensize II}]/H\alpha). \]

This kind of relation has been found by Storchi-Bergmann et al. (1994) for a sample dominated by more metal-rich starbursts. The main difference encountered here is that, for the lower metallicity H{\sevensize II} galaxies of Fig. \ref{nii}, [N{\sevensize II}]/H$\alpha$ shows a wider dynamical range as a function of $12+log$(O/H) than do the more metal-rich Starbursts.   

\subsection{Age indicators}

The equivalent width W(H$\beta$), the [O{\sevensize III}]$\lambda\lambda$4959,5007/H$\beta$ ratio and relative volume $R$ of the He$^+$ and H$^+$ zones are age indicators for H{\sevensize II} regions (Dottori, 1981, 1987). Dottori and Bica (1981) have shown that W(H$\beta$) can be used to date H{\sevensize II} regions of the Magellanic Clouds. Copetti et al. (1986) studied the evolution of the above properties with age through stellar evolution models with a single burst of star formation, which can be used to derive the age of the ionizing star clusters from W(H$\beta$) or [O{\sevensize III}]/H$\beta$.

More recently, Stasi\'{n}ska and Leitherer (1996) constructed a new grid of models representing an H{\sevensize II} region produced by an evolving starburst embedded in a gas cloud of the same metallicity. They concluded that both W(H$\beta$) and W([O{\sevensize III}]) were good age indicators, but not  [O{\sevensize III}]/H$\beta$. According to them, the [O{\sevensize III}]/H$\beta$ ratio is a poor chronometer because its time dependence is mild for ages smaller than 5 Myr and its behaviour is affected by  the ionization parameter.

\begin{figure*}
\vspace{16cm}
\begin{tabular}{cc}
\includegraphics{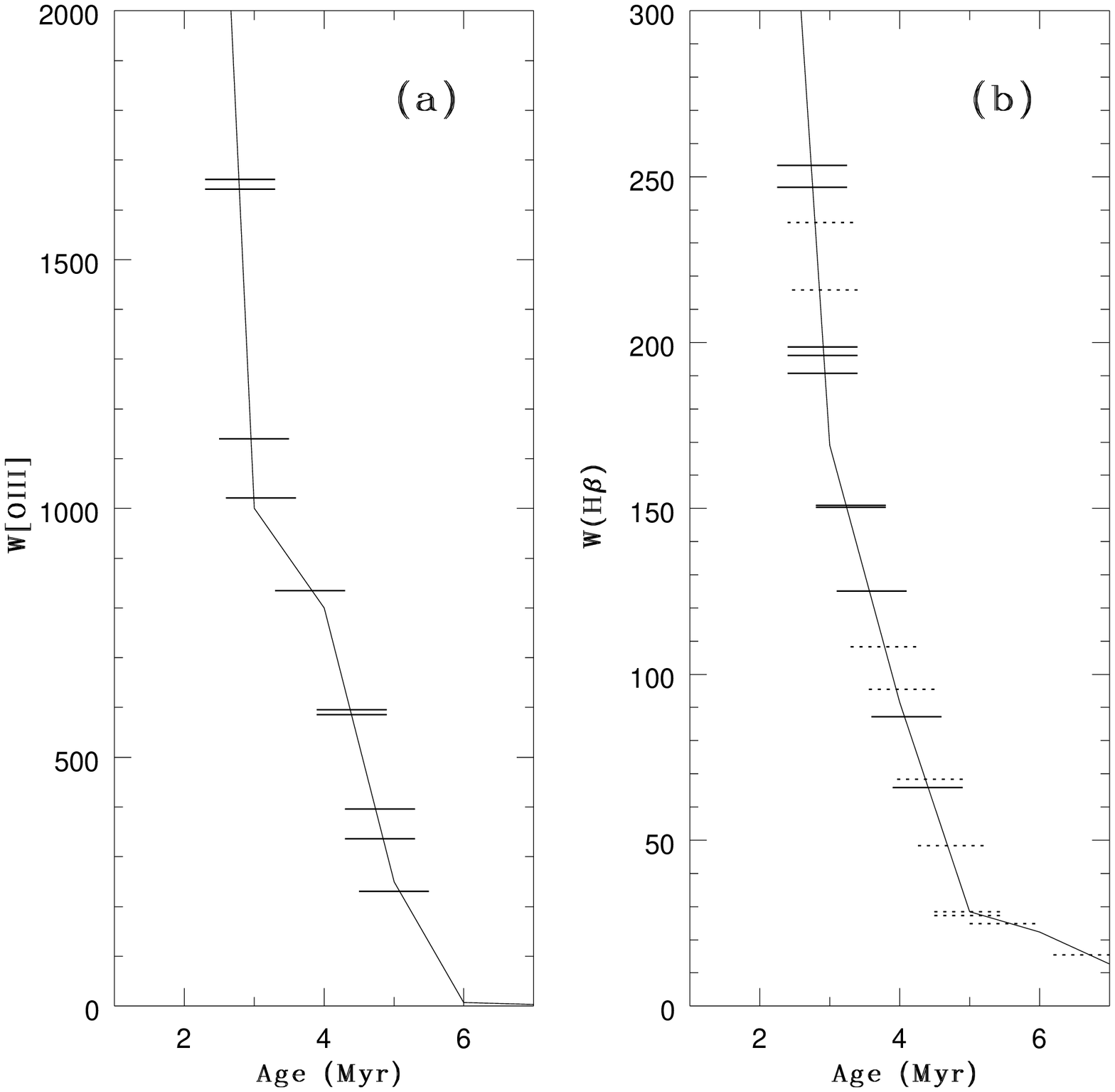} \\
\includegraphics{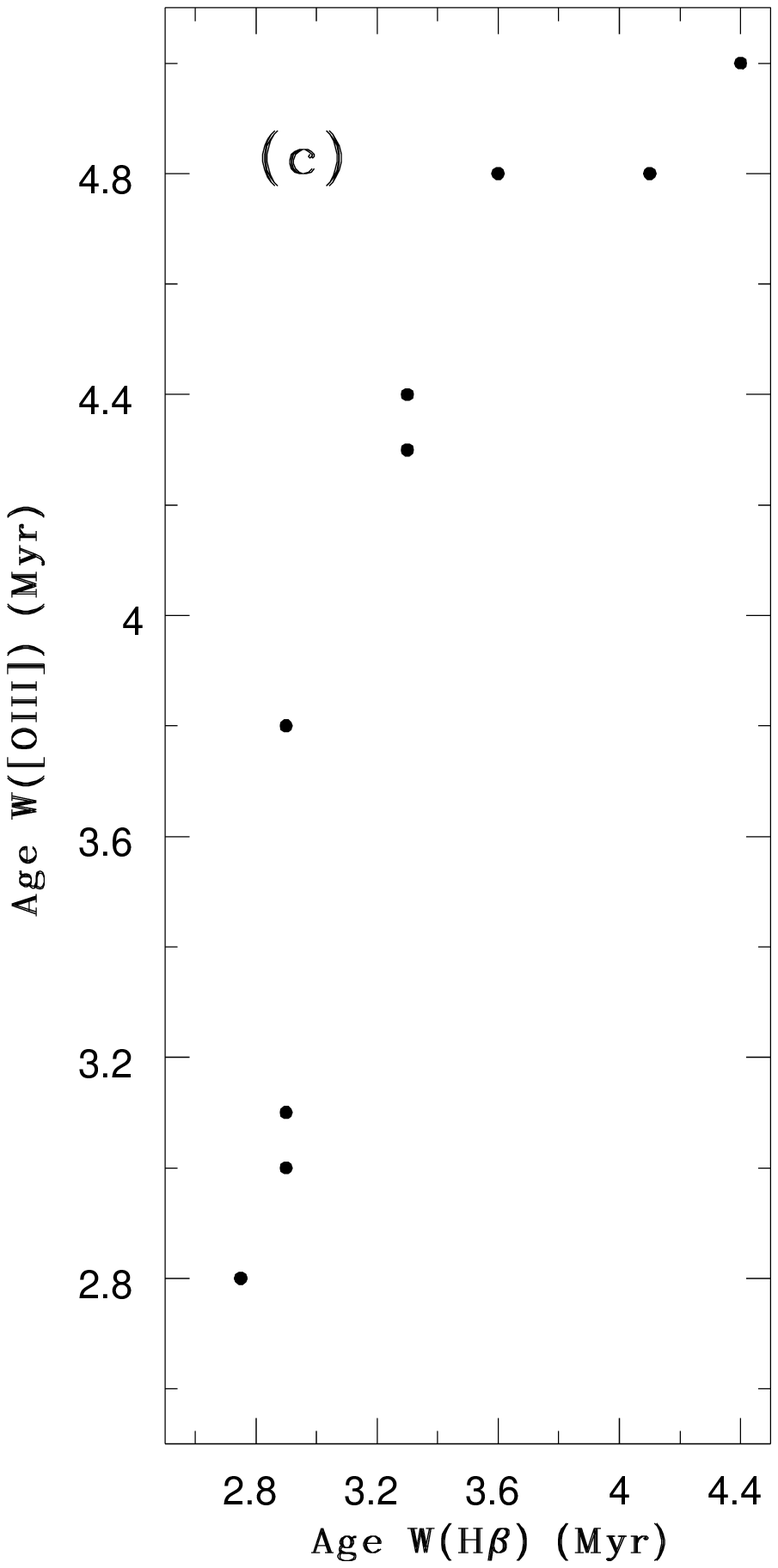}
\end{tabular}
\caption{Model sequences (continuous lines) of (a) W([O{\sevensize III}]) and (b) W(H$\beta$) as a function of age of the ionizing stellar population (Stasi\'{n}ska and Leitherer, 1996) together with the values of W([O{\sevensize III}]) and W(H$\beta$) for the present sample (horizontal solid lines) used to derive the ages of ionizing stellar population. In (b) we also show the data previously to the stellar population subtraction (horizontal dashed lines). The age derived with W([O{\sevensize III}]) is plotted against the age derived with W(H$\beta$) in (c).}
\label{Age1}
\end{figure*}

We have thus used the models of Stasi\'{n}ska and Leitherer (1996) to derive the age of the ionizing star cluster for the present H{\sevensize II} galaxy groups from the measured W({H$\beta$}) and W([O{\sevensize III}]). In Figs. \ref{Age1}a and \ref{Age1}b we present a sequence of models (continuous line) in W([O{\sevensize III}]) and W(H$\beta$) versus age diagrams together with the data from our groups (horizontal lines). The models with the parameters best suited to our data correspond to $Z=0.25Z_{\odot}$.

In the above models the continuum is only due to the nebular and ionizing stellar population. As concluded in Paper I, in the galaxy groups there is additional contribution of older stellar populations to the continuum, so that it is necessary to subtract that contribution before calculating W(H$\beta$) and W([O{\sevensize III}]). In order to illustrate the effect of this additional component in the derived ages, we present in Fig. \ref{Age1}b both the data previously to the subtraction as dashed lines, and the corrected data as solid lines. 

The age derived from W(H$\beta$) is plotted against the age derived from W([O{\sevensize III}]) in Fig. \ref{Age1}c. Although the ages derived from W([O{\sevensize III}]) are in most cases up to $\approx$ 1 Myr larger than the ones from W(H$\beta$), they are well correlated with Spearman rank correlation cofficient of 0.98. Using both diagrams (Figs. \ref{Age1}a and \ref{Age1}b) we derive an age range of 2.7--5.0 Myr for the ionizing populations in the H{\sevensize II} galaxy groups. This age range corresponds to the predicted Wolf-Rayet (WR) phase of a young star cluster (Schaerer \& Vacca, 1998). Indeed,  WR features have been detected in our H{\sevensize II} galaxy spectra and were discussed in Paper I. 

As the age does not vary much among the H{\sevensize II} galaxy groups, the sequence defined in the BPT diagram must be dominated by a varying metallicity (Sect. 4.1).

\begin{figure}
\vspace{8cm}
\includegraphics{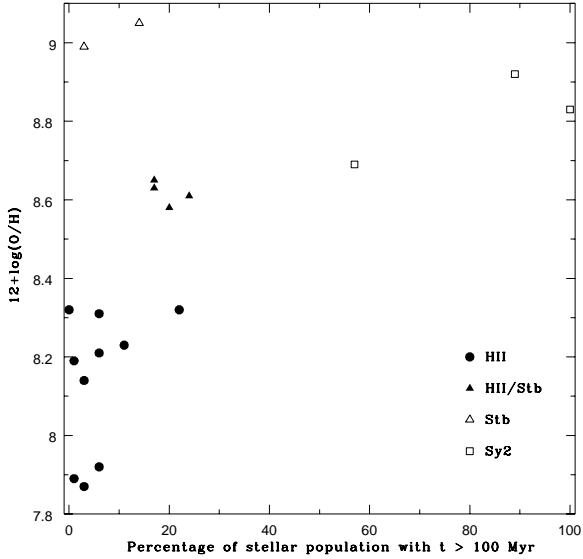}
\caption{Relation between metallicity and contribution of stellar populations (to the flux at $\lambda = 4020$ \AA) older than 100 Myr to the spectrum. Circles represent H{\sevensize II} galaxies, squares Seyfert's, open triangles Starbursts and filled triangles Intermediate H{\sevensize II}/Starbursts.}
\label{met1}
\end{figure}

An age effect is suggested when we investigate the relation between the gas metallicity and the contribution of the stellar population with ages $t > 100$ Myr to the spectra. This can be observed in Fig. \ref{met1} where we plot $12+log$(O/H) against the percentage of stellar population with ages $>$ 100 Myr. This diagram suggests that most of the gas enrichment has been produced by stellar populations older than 100 Myr. The H{\sevensize II} galaxies are the ones with smallest contributions of stars older than 100 Myr and most metal-poor while the Seyfert's have the largest contributions of older populations and are the most metal-rich. There are only two deviant points, corresponding to the Starbursts. These cases can be understood as due to exceptionally high star formation rates in luminous evolved galaxies, spiral or interacting, so that light from stars younger than 100 Myr dominate the observed fluxes of the galaxies.

\section{Possible cosmological uses}

Recently, Melnick et al. (1999) have called attention to the possible use of H{\sevensize II} galaxies as cosmological probes. In principle (i) emission line velocity dispersions, (ii) H{\sevensize II} galaxy luminosities as candles, and (iii) stellar population/emission line properties as probes of galaxy evolution would be important tools to explore. In the following we discuss the latter two possibilities.   

As can be seen in Table \ref{grupos} the $<$M$_B$$>$ values vary a lot among H{\sevensize II} galaxy groups. Although the sources of B magnitudes are heterogeneous we do not expect that broad band magnitudes be useful as candles, since the contribution from underlying stellar populations of different ages varies considerably, as found in the population syntheses (Paper I). However, H$\beta$ emission line luminosities appear to be less dispersed both within and among groups. This is especially so, by considering aperture effects among groups (Table \ref{grupos}). In order to further explore this possibility it would be important to observe whole sets of local calibrators by means of narrow filter CCD imaging in emission lines and neighbouring continuum. 

For high redshift galaxies, many lines used in the traditional diagnostic diagrams move out of the optical spectral range and typically the spectra have a low S/N ratio in the continuum. Frequently, in these cases, the two strongest emission lines observable in the optical window are [O{\sevensize II}]$\lambda3727$ \AA\ and H$\beta$. Rola et al. (1997, hereafter R97) used the equivalent widths of these lines to construct new diagnostic diagrams in order to classify the spectra of distant emission-line galaxies observed in deep galaxy redshift surveys.

\begin{figure}
\vspace{8cm}
\includegraphics{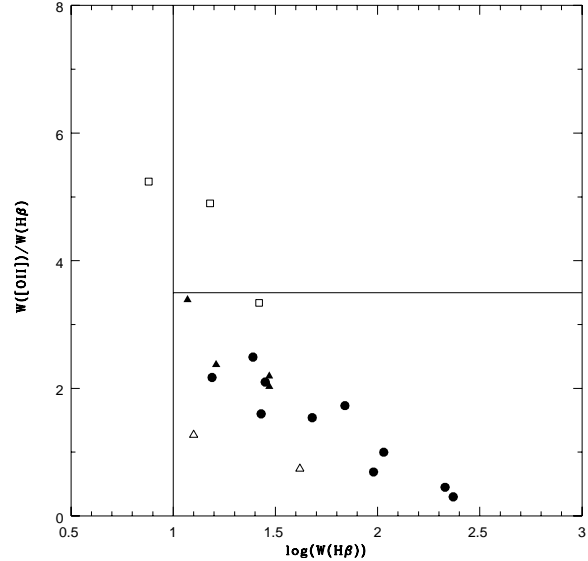}
\caption{Diagnostic diagram W([O{\sevensize II}])/W(H$\beta)  \times log($W(H$\beta))$ adapted from Rola et al. (1997) with our data without subtracting the contribution of older stellar populations to the continuum. Symbols as in Fig. \ref{met1}.}
\label{rola1}
\end{figure}

Our data in  the  diagnostic  diagram  W([O{\sevensize II}])/W(H$\beta)  \times log($W(H$\beta))$ are shown in Fig. \ref{rola1}. In agreement with R97, the Seyfert galaxies occupy the region with W([O{\sevensize II}])/W(H$\beta) \geq 3.5$ while the Starbursts and H{\sevensize II} galaxies occupy the region with W([O{\sevensize II}])/W(H$\beta) \leq 3.5$ and $log($W(H$\beta))>1.0$. Although this diagram segregates AGNs from H{\sevensize II}/Starburst galaxies we find that it does not segregate between H{\sevensize II} and Starburst galaxies. 

\begin{figure}
\vspace{8cm}
\includegraphics{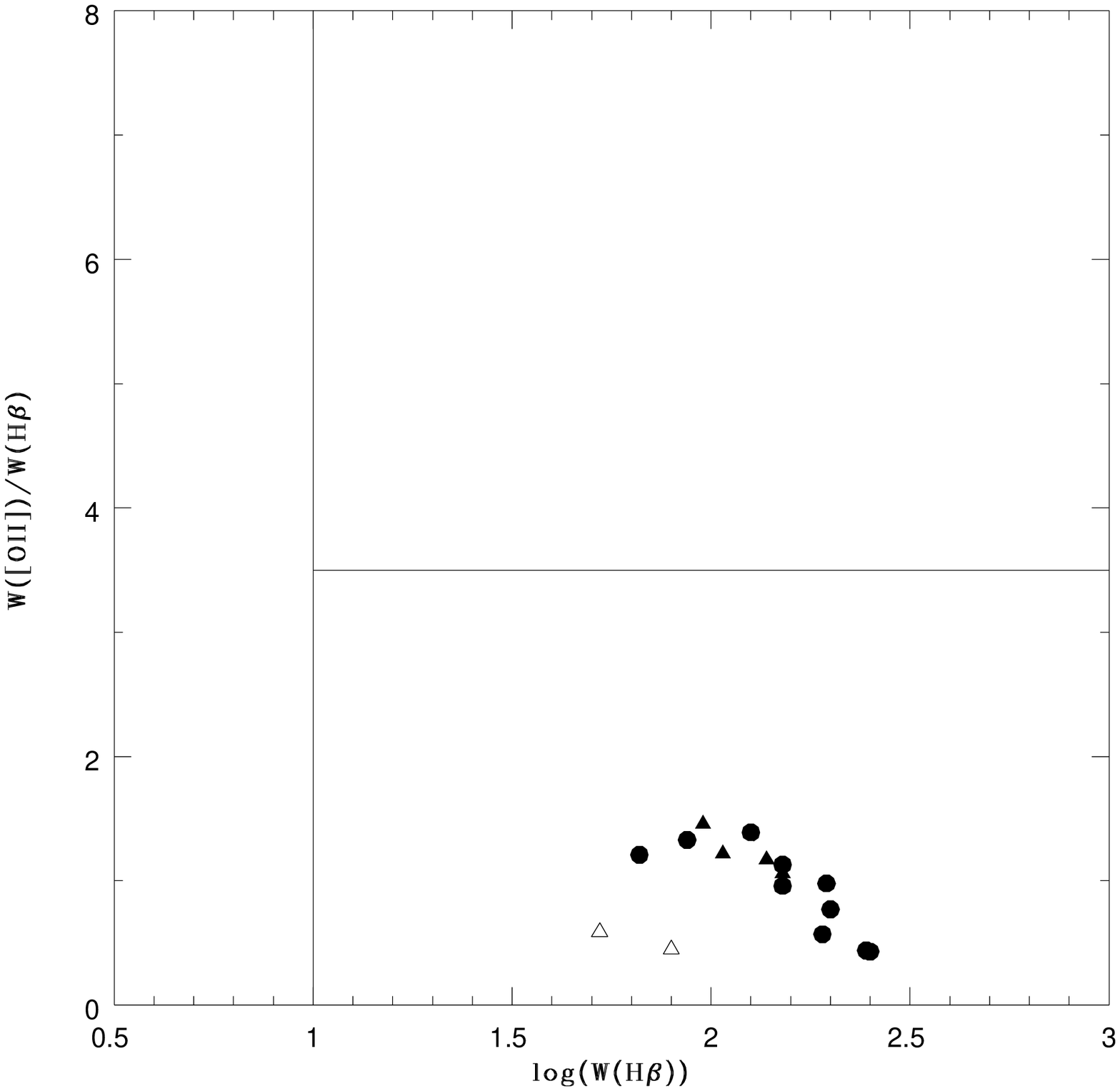}
\caption{Diagnostic diagram W([O{\sevensize II}])/W(H$\beta)  \times log($W(H$\beta))$ adapted from Rola et al. (1997) with our data after subtracting the contribution of older stellar populations to the continuum. Symbols as in Fig. \ref{met1}.}
\label{rola2}
\end{figure}

\begin{figure}
\vspace{8cm}
\includegraphics{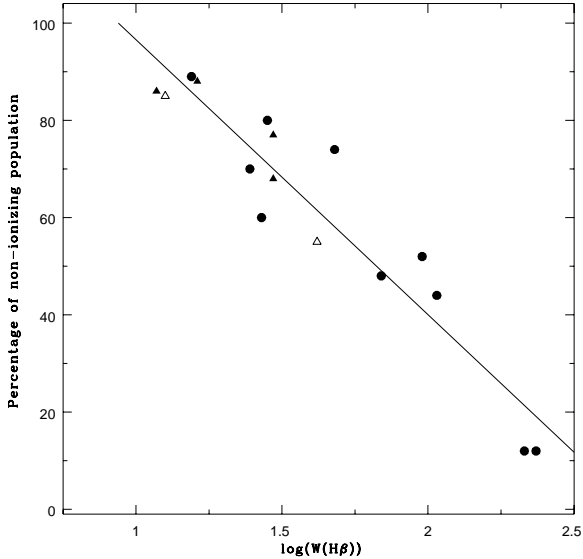}
\caption{Relation between percentage of non-ionizing stellar population to the flux at $\lambda=$ 4020 \AA\ and $log($W(H$\beta))$. The straight line is a linear regression to the data. Simbols as in Fig. \ref{met1}.}
\label{non1}
\end{figure}

We have investigated the effect of the stellar population contribution in the above diagram, by subtracting the contribution of the older age components to the continuum of the H{\sevensize II} and Starburst galaxies, and leaving only the ionizing continuum for the calculation of the equivalent widths. The result is presented in Fig. \ref{rola2} which shows a shift in the loci of the galaxies with W(H$\beta)$ as low as $\approx$ 10 \AA\  to values larger than 50 \AA, and for the W([O{\sevensize II}])/W($H\beta)$ from values as large as $\approx 3$ down to values lower than 1.5. It can be concluded that equivalent widths W([O{\sevensize II}]) and W(H$\beta)$ of Starburst and H{\sevensize II} galaxies show a much smaller range of values once the older stellar population contributions are subtracted, which means that the intrinsic properties of star forming regions do not vary as much as suggested by the diagram before subtraction (Fig. \ref{rola1}). In fact, we conclude that the value of $log($W(H$\beta))$ can be used as an indicator of the percentage contribution of the non-ionizing stellar population to the continuum. In Fig. \ref{non1} we plot this percentage $P$ against $log($W(H$\beta))$ together with a linear regression to the data: 

\[ P(\%)=153.04(\pm8.76)-56.52(\pm5.31) \times log(W(H\beta)), \]

\noindent for W(H$\beta)$ in \AA.

The Spearman rank correlation coefficient to this correlation is -0.90. W(H$\beta)$ is thus a powerful tool for estimating the contribution of older stars to the spectrum of a galaxy. The relation above is very useful, especially for distant galaxies, for which W(H$\beta)$ can be obtained once a continuum can be detected underneath H$\beta$. 

On the other hand, back to the R97 diagram, we can conclude that the correction by the stellar population does not move the H{\sevensize II} and Starburst galaxies to regions in the diagram outside the limits suggested by R97. But, for very distant galaxies, age ranges will necessarily become narrower and fractions of intrinsically young galaxies should increase; in both cases the lower right part of R97's diagram should become more and more populated for such high redshift samples, as simulated in our analysis above.

\section{Concluding remarks}

We have analysed the emission-line spectra of 19 galaxy templates, obtained from the grouping of 185 emission-line galaxy spectra, as described in Paper I. After correction for internal reddening and subtraction of the synthesized population spectrum, both from Paper I, the emission-line fluxes of each template were measured and corrected for additional gaseous reddening, when present. According to the corresponding locus of each template in the BPT diagrams, the present sample comprises 10 groups of H{\sevensize II} galaxies, 3 groups of Seyfert 2, 2 groups of Nuclear Starbursts and 4 groups of intermediate cases between Nuclear Starbursts and H{\sevensize II} galaxies. 

The H{\sevensize II} galaxy groups define a much tighter sequence in the BPT diagrams than the individual galaxies, suggesting that the spread obtained in previous studies is due to a combination of lower S/N ratio spectra and of not taking into account the contribution of the underlying stellar population. With population subtraction in spectra of improved S/N ratio, the H{\sevensize II} galaxy groups get closer to the theoretical sequences of H{\sevensize II} regions presented by Evans and Dopita (1985) and McCall et al. (1985), suggesting that they are more similar to H{\sevensize II} regions than concluded in previous works. The resulting sequence of H{\sevensize II} templates in the BPT diagram is suggested as a fiducial observational locus for future models of H{\sevensize II} galaxies.

From the emission-line ratios, we calculate the gas metallicity and age of the ionizing stellar population, and investigate the effect of these two parameters in the BPT diagram above. We conclude that the sequence defined
by the H{\sevensize II} galaxy templates is primarily due to metallicity, which spans the range 7.87 $<$ 12+$log$(O/H) $<$ 8.32 ($\approx$ 1/11 to 1/4 solar), while the age of the ionizing stellar population ranges only from 2.7 to 5.0 Myr.

A connection with age is suggested when we relate the gas metallicity with the percentage contribution of stellar population components older than 100 Myr, which may indicate that the metal enrichment is mostly due to previous stellar generations, whose signatures are present in the spectral distribution even for the bluest H{\sevensize II} galaxies, as discussed in Paper I. The larger the contribution of older stellar components, the more metal rich is the gas. We find a good correlation between $log($[N{\sevensize II}]/H$\alpha)$ and $12+log$(O/H) and propose a calibration to obtain the latter from the former.

We also explore the effect of the stellar population contribution to the equivalent width diagnostic diagrams of Rola et al. (1997). We conclude that the observed ranges in W([O{\sevensize II}])/W(H$\beta)$ and W(H$\beta)$ are essentially due to the non-ionizing stellar population contribution. By relating $log($W(H$\beta))$ to the percentage contribution of this population, we conclude that there is a tight correlation between these two quantities. We thus propose a calibration which can be used to estimate the  non-ionizing stellar population contribution to the spectra from the measured W(H$\beta)$, which is particularly useful for probing properties of distant galaxies.

\section*{Acknowledgments}
T.S.B., E.B. and H.S. (during part of this work) acknowledge support from the Brazilian Institution CNPq, and D.R. from CAPES. We thank Iranderly F. de Fernandes (as CNPq undergraduate fellow) for work related to this project. We thank an anonymous referee for valuable suggestions which helped to improve the paper.

\label{lastpage}

\end{document}